# Art Pricing with Computer Graphic Techniques

Lan Ju

Associate Professor in Finance

Peking University HSBC Business School

Zhiyong Tu

(Corresponding Author)

Associate Professor in Economics

Peking University HSBC Business School

zytu@phbs.pku.edu.cn

Changyong Xue

Ph.D. in Economics

Peking University HSBC Business School

September 2019

ABSRACT

This paper makes the first attempt to introduce the tools from computer graphics into the art pricing research. We argue that the creation of a painting calls for a combination of conceptual effort and painting effort from the artist. However, as the important price determinants, both efforts are long missing in the traditional hedonic model because they are hard to measure. This paper draws on the digital pictures of auctioned paintings from various renowned artists, and applies the image recognition techniques to measure the variances of lines and colors of these paintings. We then use them as the proxies for the artist's painting effort, and include them in the hedonic regression to test their significance. Our results show that the variances of lines and colors of a painting can significantly positively explain the sales price in a general context. Our suggested measurements can better capture the content heterogeneity of paintings hence improving on the traditional art pricing methodology. Our approach also provides a quantitative perspective for both valuation and authentication of paintings.

## 1. Introduction

Computer graphics is a discipline that studies the representation and manipulation of image data via computer algorithms. We think it is natural to introduce the techniques from computer graphics to investigate the pricing of artworks, especially paintings, since we can now easily access their digital images in most major databases. This paper makes the first attempt to retrieve more concrete information of a painting's content by handling its image data, and see if it can improve those traditional art pricing models. Our main premise is that a higher level of information in a painting (defined later) tends to imply a more intense artistic effort during its creation; hence it shall be priced higher.

In the art literature, the sales price of a painting is often explained by the multiple factors ranging from painting attributes such as size, material and signature to sales conditions such as year, salesroom and sales location, etc. This approach is named as the hedonic model, which is extensively applied in the analysis of art investment (e.g., Buelens and Ginsburgh, 1993, Chanel, 1995 and Taylor and Coleman, 2011, among others).

Within those aforementioned price determinants, the size of the painting is normally found to be a significant explanatory variable to price. For example, Etro and Stepanova (2017) studied a sales sample around 200 years ago (1780-1840) and revealed a significantly positive size effect and a negative effect of the square of size to the auction price for the early art market. Anderson et al. (2016) adopted a more recent sample (1987-2011) to examine the investment properties of American art. Similarly, they showed that the coefficient of size is significantly positive and that of the quadratic term of size is significantly negative, which demonstrates a decreasing marginal effect of size to the sales price.

This type of size effect is quite intuitive. The larger is a painting, the more effort is needed by the artist to produce, so the higher it shall be priced. Also, the price is not linearly related to the size, because not only the size but also the content of the painting plays a vital role in determining its sales value. To control for the content heterogeneity, many papers introduce the content dummies in the hedonic model. For example, Renneboog and Spaenjers (2013) categorized their sample into eleven groups by the distinct subject matters, and accounted for their influence on the sales price using Topic dummies.[1]

This paper argues that the usual topic dummies only give a rather limited description of the painting content. Within the same category, the painting contents, consequently their sales prices can vary greatly. For two equally sized paintings with the same subject matter, the one with just a few sketches must be easier to paint than the one with a more sophisticated composition, so the latter one shall be price higher. The following is a real example.

[1] The eleven categories are: ABSTRACT, ANIMALS, LANDSCAPE, NUDE, PEOPLE, PORTRAIT, RELIGION, SELF-PORTRAIT, STILL_LIFE, UNTITLED, AND URBAN.

We notice that two Picasso's paintings, "*Homme À La Pipe Et Nu Couché, 1967*" and "*Nu Couché, 1967*", have the same physical features (length 146cm, width 114cm and oil canvas). They were sold in the same year of 2011 in the Sotheby's, London, but the former one obtained 4,801,250 GBP, two times that of the latter (2,281,250 GBP). Both paintings were created in 1967 hence belonging to the same period of Picasso. They also depict the same subject matter. The main difference is that the former contains more human figures than the latter, so it needs more painting effort to produce.

Most literature in art investment, such as Biey and Zanola (2005), Lazzaro (2006) and Ginsburgh et al. (2019) etc., just use the conventional dummy variables to control for the content heterogeneity (e.g., sceneries, figures, among others). From the above example, we can see that these dummies may be too coarse to effectively differentiate artworks. It demonstrates that we are lack of a more elaborate measurement for the content details of a painting. The main contribution of this paper is to suggest a pair of objective quantitative measurements for the painting's content based on the tools from computer graphics, and show that they can significantly explain the sales price of the painting in a very general sense.

We hold that the creation of a painting requires a combination of two types of efforts from the artist: the conceptual effort and the painting effort. The latter effort aims to build a physical representation of the former. Little literature has tried to account for these two different types of efforts, probably because they are hard to measure. Drawing on the computer graphic techniques, we try to quantify the artist's painting effort that is crystalized in a painting's content. While for the conceptual effort, it will still be hiding in the regression residue as usual.

Notice that a more delicate painting (such as Picture (A) in Figure 2 in Section 2) generally requires more effort to paint once it has been conceptualized. By this logic, we could use the degree of the painting delicacy as a proxy for the artist's painting effort for a given painting size. Then how to measure the painting delicacy? We argue that since the basic building blocks of a painting are lines and colors, we can define the degree of delicacy as the degree of variations of lines and colors on the painting. Figure 1 in the next section gives a more vivid explanation for such constructions.

Based on the digital image of a particular painting, we can calculate the variances of lines and colors that comprise of the painting, and use them as the measurements for the artist's painting effort.[2] After obtaining these variances, we could test how they would influence the sales price of the painting via a hedonic regression model. Our benchmark analysis mainly draws on the sales records of the paintings by Pablo Picasso. These paintings were auctioned between the years of 2000 to 2018.

Picasso's artworks are a constant research topic in the art investment literature (e.g., Czujack et al., 2004, Forsund et al., 2006, Pesando and Shum, 2007 and Scorcu and Zanola, 2011). One reason is that Picasso is both prolific and revolutionary; his paintings in the market not only constitute a large sample, but also range across diverse styles. Another reason may be that Picasso's prints are actively traded and fairly liquid so the formation of their prices is highly marketized (Pesando and Shum, 1999 and Biey and Zanola, 2005).

Our main finding based on the sample of Picasso is that both the variances of lines and colors are significantly positive to explain the sales price of the painting. Our robustness tests extend the analysis to the other artists, e.g., the French impressionism artist Pierre-Auguste Renoir and the Chinese traditional painting artist Baishi Qi. We find our results are fairly robust: the complexity/delicacy of a painting in terms of the variances of its lines and colors is significantly positively related to the painting's price.

To our knowledge, this paper is the first to introduce the image recognition technique into the artwork pricing research. It reinstates the long missing variables for the traditional hedonic model, which can better explain the heterogeneity of painting contents. Even for the repeated sales model where only the items sold at least twice are considered (e.g., Goetzmann, 1993, Pesando, 1993, Mei and Mose, 2002 and Park et al., 2016), our measurements can also serve as the new controls in order to obtain a more accurate inference.

From a practical point of view, our method can help to identify the different contributions of the painting effort versus conceptual effort to the artwork's aggregate price formation. The disentanglement of the price influences of these two efforts provides a valuable guidance for the painting's art history positioning. In addition, our approach may also provide a quantitative perspective on both valuation and authentication of paintings.

The reminder of the paper is organized as follows. Section 2 elaborates on the proposed measurements of the variations of lines and colors of a painting. Section 3 carries out the benchmark regression based on the Picasso sample. Section 4 applies our new measurements to various contexts so as to test the robustness of their efficacy. Section 5 concludes.

[2] Please see the Appendix A for the detailed procedure to generate these variances based on the digital picture.

## 2. Painting effort versus conceptual effort

We first clarify our classification of painting effort versus conceptual effort during the artist's creation of a painting. Then we explain why the variances of lines and colors are reasonable measurements for the artist's painting effort.

### *2.1 The creation of a painting*

A painting is a form of crystallization of the artistic effort. The artist will compose a painting based upon its building blocks, i.e., lines and colors. The infinitely possible combinations of lines and colors provide an unlimited scope for creative concepts, where a concept dictates a particular kind of arrangement for lines and colors aiming at the transmission of certain meanings/feelings.

For example, the innovative Cubism concept at the beginning of the twentieth century breaks up the objects and reassembles in an abstracted form. Contrary to the traditional concept to depict objects from a single viewpoint, Cubism depicts the subject from a multitude of viewpoints so as to represent a physical and psychological sense of the fluidity of consciousness. Developing such concept calls for the conceptual effort from the artist.

Then, the artist will convey her concept/notion to the public through the physical media—the painting. The painting must be crafted by the artist with the relevant techniques using brushes as well as applying colors. We call this painting effort. Obviously, the creation of a painting requires both the conceptual and the painting efforts.

The art literature also provides the evidence for such kind of separation of conceptual versus painting efforts during the production of artworks. For example, Galenson (2009) documented that many renowned artists, such as Rubens, Moholy-Nagy and even Picasso, often ask their assistants to complete the artworks for them after giving their conceptual indications.[3]

For most hedonic models, the effects of conceptual and painting efforts on prices are hard to separate since they are normally aggregated into the error term. This paper tries to measure the painting effort quantitatively. By extracting the painting effort/value from the overall price, people can more accurately judge the different contributions of each effort (painting versus conceptual) on the valuation of a particular artwork. Currently such tasks have to rely on the experts' subjective opinions.

[3] Painting by Proxy: The Conceptual Artist as Manufacturer, David W. Galenson, 184-198. In Conceptual Revolutions in Twentieth-Century Art, Galenson. 2009

*2.2 Measurements of the painting effort*

We will define the variances of lines and colors of a painting, and use them to quantify the variations of these building blocks of the painting. We hold that they can be good proxies for the painting effort of the artist during her production of the artwork.

The following Figure 1 gives a heuristic explanation for the proposed variance measures. The idea is to look at a digital picture at its micro level, i.e., in the form of its basic elements of pixels. At such a level, the lines and colors become dots covering each segment of pixel, and then we can evaluate the variation of these segmental dots.

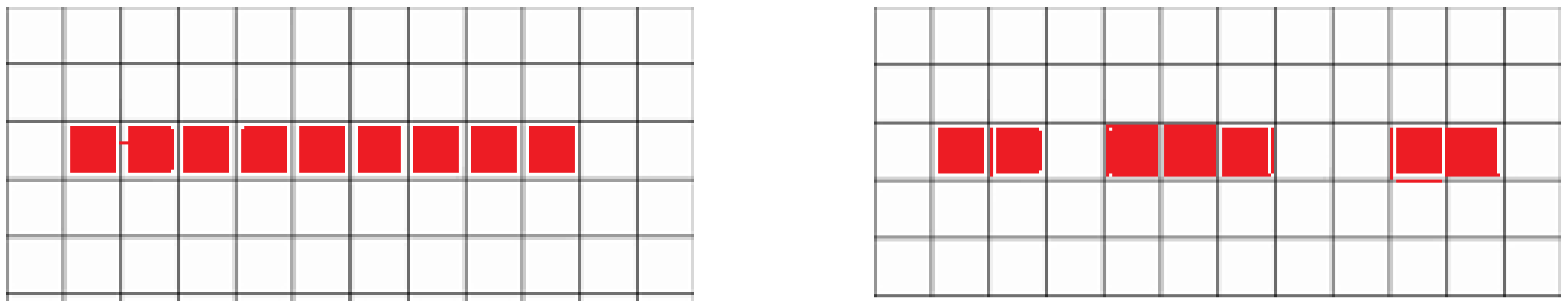

**(A)**

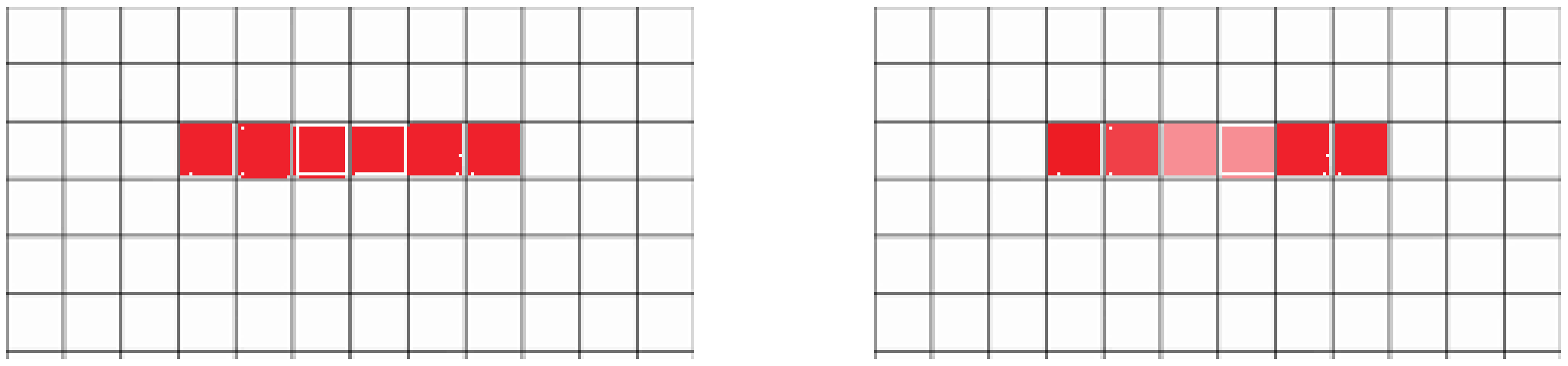

**(B)**

**Figure 1. Representation of the Variations of Lines and Colors at the Pixel Level**

In Figure 1 (A), a line crosses nine consecutive pixels in the left picture; while in the right picture it takes a break after crossing the second pixel, and takes two breaks after crossing the sixth. Obviously, the variation of lines in the right picture is larger. To produce this variation, the initial mental mode of drawing a line must be switched to the mode of a blank, and then resumes to the line mode again. This kind of phase-switching triggers more intense brain activity than just maintaining a constant mental state (i.e., a continuous line).

The above argument can find support from the neuroscientific literature. For example, both Matsuda et al. (2017) and Yokoyama et al. (2018) show that the alpha-power inter-parietal synchronization will be enhanced only when the selected action is switched from a previous action to a different action. So a larger variance of lines in a painting implies more frequent mental phase switching when producing the whole picture, which naturally requires more effort/energy from the artist.

Similar logic/mechanism applies to Figure 1 (B), which depicts different degrees of color variations for the same straight line. The line color in the left picture is a pure red, while in the right an asymptotic red. So the line in the right picture has a larger variance in colors. Consequently, generating the right line will involve more frequent mental state switching, hence more energy.[4]

The following Figure 2 provides an illustrative example of two pictures with apparently different delicacies, i.e., different variances of lines and colors. Based on the variance formula in Appendix A, we can calculate that the variance of lines for the left rooster is two times that of the right. The variance of colors for the left rooster is thirty percent higher than that of the right. So by our measurements, we can claim that the left rooster will require much more painting effort than the right one.

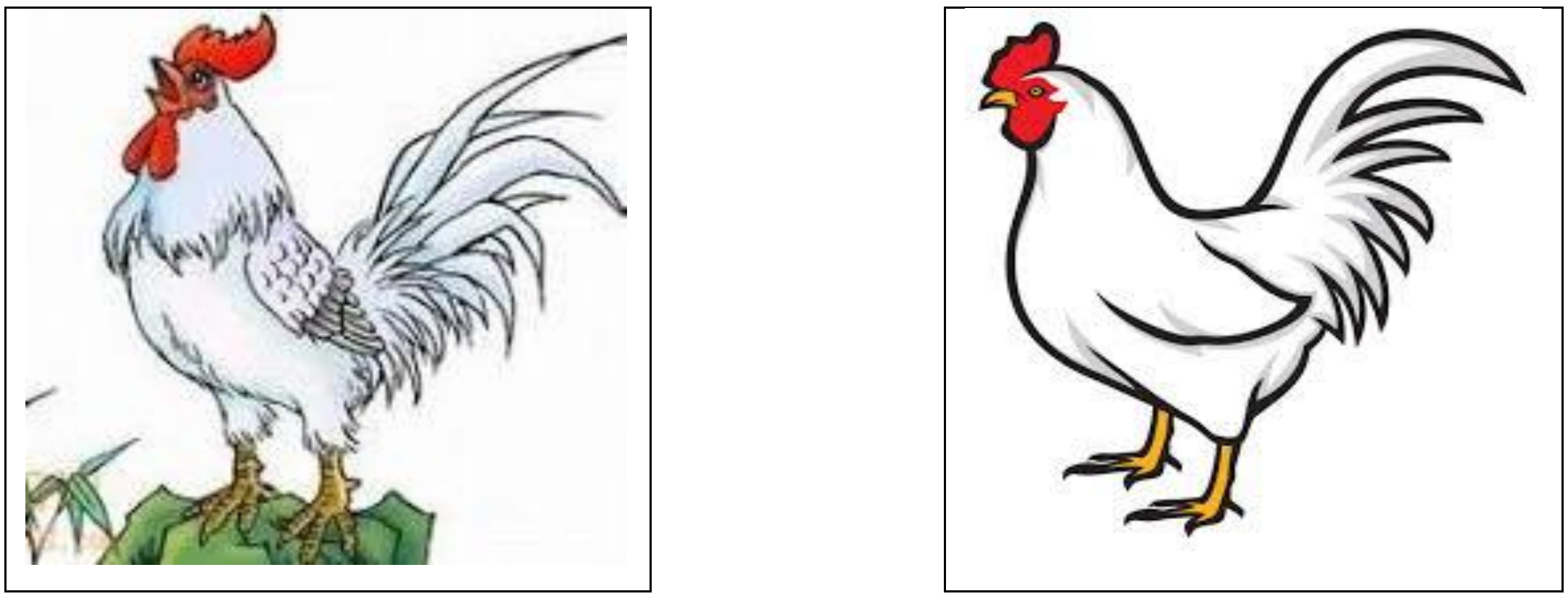

(A) (B)

**Figure 2. Two Roosters with Different Delicacies (Variances of Lines and Colors)**

Notice that painting effort can also be interpreted as a kind of shadow price of replication. More sophisticated artworks build higher barriers for imitation (because they need more effort to produce), hence may be priced higher in the market after controlling for the other factors.

## 3. The benchmark regression

Our benchmark regression focuses on the paintings by Pablo Picasso that span across all of his artistic stages. Our results show that both the variances of lines and colors of the painting are significantly positively related to the painting's auction price.

[4] This argument is true for a given artist endowed with a given level of craftsmanship. But for different artists with heterogeneous skills, the same variations of lines and colors may cost different efforts. In such contexts, we can interpret our proposed variances as a kind of *standardized* measurements of painting efforts.

*3.1 Data and methodology*

The data are drawn from the Blouin Art Sales Index (BASI), presently the largest known database of artworks online that provides data on artworks sold at auction at over 350 auction houses worldwide (Korteweg et al., 2015). We adopt all the records of Picasso's paintings from year 2000 to 2018 except those without pictures or prices.[5] Each sales record includes such information as the title of the painting, digital image, size, date of creation, materials, date and city of sales, salesrooms, signature and prices, etc.[6] Altogether, we obtain 720 records.

For the robustness analysis, we draw on the auction records of the same period from another two famous artists. One is the French artist Pierre-Auguste Renoir (1841-1919), whose data are also from BASI. The other is the Chinese artist Baishi Qi (1864-1957), and his data are from Artron, from which we exclude those auctioned in mainland China.[7] We wish to further test whether our results hold for artists across different countries and styles.

The following is our benchmark regression of the hedonic model:

$$p_{it} = \theta_1 A_i + \theta_2 B_i + \sum_{k=1}^{m} \alpha_k x_{ki} + \sum_{t=t_0}^{T} \varphi_t D_{it} + \varepsilon_i \quad (1)$$

The regression includes the usual explanatory variables related to the painting, and our innovation here is to introduce two measurements of painting effort, $A_i$ and $B_i$, into the model. $A_i$ and $B_i$ calculate the variances of lines and colors for a given painting respectively (see Appendix A for their detailed constructions). $p_{it}$ is the price of painting *i* sold as time *t*, $x_{ki}$ is defined as the set of time-invariant characteristics of painting $i$ (e.g., the size, material and signature). $D_{it}$ is the set of time varying idiosyncratic attributes (e.g., year dummy) for painting $i$.

*3.2 Descriptive statistics*

Based on the auction data of Picasso from year 2000 to 2018, we obtain the following summary statistics for all the regression variables in Table 1.

[5] Note that the records before 1997 in BASI do not provide the digital pictures, and the auction records between 1997 and 1999 are very sparse.

[6] Note that all the hammer prices are converted into U.S. dollars at the spot exchange rates at the time of sales.

[7] The records of Baishi Qi are incomplete in BASI. So we refer to Artron, which is the biggest professional art information release carrier in China (www.Artron.net). To avoid the geographic heterogeneity, we exclude those auction records in mainland China.

**Table 1. Summary Statistics of the Regression Variables**

| VARIABLES | N | Mean | Sd | Min | Max |
|---|---|---|---|---|---|
| Price($) | 720 | 5,893,419 | 1.20e+07 | 1,280 | 179,365,000 |
| Line | 720 | 0.093 | 0.019 | 0.040 | 0.156 |
| Color | 720 | 0.213 | 0.086 | 0.006 | 0.441 |
| Age | 720 | 74.68 | 20.63 | 47 | 125 |
| Salesyear | 720 | 2,010 | 5.203 | 2,000 | 2,018 |
| Surface(1000cm$^2$) | 720 | 6.378 | 10.67 | 0.0160 | 163.5 |
| Signature | 720 | 0.564 | 0.496 | 0 | 1 |
| Dated | 720 | 0.617 | 0.487 | 0 | 1 |
| Material | | | | | |
| board | 720 | 0.047 | 0.212 | 0 | 1 |
| burlap | 720 | 0.061 | 0.240 | 0 | 1 |
| canvas | 720 | 0.815 | 0.388 | 0 | 1 |
| cardboard | 720 | 0.022 | 0.147 | 0 | 1 |
| ceramic | 720 | 0.033 | 0.180 | 0 | 1 |
| others | 720 | 0.021 | 0.143 | 0 | 1 |
| City | | | | | |
| London | 720 | 0.358 | 0.480 | 0 | 1 |
| New York | 720 | 0.549 | 0.498 | 0 | 1 |
| Paris | 720 | 0.063 | 0.242 | 0 | 1 |
| Others | 720 | 0.031 | 0.172 | 0 | 1 |
| Salesroom | | | | | |
| Christie's | 720 | 0.518 | 0.500 | 0 | 1 |
| Sotheby's | 720 | 0.428 | 0.495 | 0 | 1 |
| Others | 720 | 0.054 | 0.226 | 0 | 1 |

In Table 1, Price is the hammer price of the painting sold in the auction denominated in USD. Line and Color are the variances of lines and colors of the painting. Age is the number of years between a painting's time of creation and its time of sale. Salesyear is the year when the painting was sold. Surface is the area of the painting, and the two dummy variables, Signature and Dated, represent if the painting is signed or dated by the artist. Other control dummies include Material (painting material), City (auction location) and Salesroom (auction house). We can see from the above summary statistics that most of the Picasso paintings in our sample are on canvas, sold by either Chrisite's or Sotheby's in London and New York.

*3.3 Regression results*

In order to smooth the data, we further logarize Price (Lprice), Line (Lline) and Color (Lcolor).[8] Then we run the cross-sectional regressions with robust standard errors for different model specifications as follows. The results are provided in Table 2.

[8] As both values of Line and Color belong to (0, 1), we inflate their values by 1000 before the logarization.

**Table 2. Results of Benchmark Regressions**

| VARIABLES | (1) Lprice | (2) Lprice | (3 Lprice | (4) Lprice | (5) Lprice | (6) Lprice |
|---|---|---|---|---|---|---|
| *Effort* | | | | | | |
| Lline | | 1.142*** | 0.537** | 42.34*** | 23.55*** | 23.89*** |
| | | (0.304) | (0.264) | (7.857) | (6.671) | (6.643) |
| $Lline^2$ | | | | -4.606*** | -2.570*** | -2.607*** |
| | | | | (0.879) | (0.738) | (0.735) |
| Lcolor | | 0.624*** | 0.404*** | -0.246 | -0.354 | 0.383*** |
| | | (0.120) | (0.0775) | (1.251) | (0.529) | (0.0783) |
| $Lcolor^2$ | | | | 0.0859 | 0.0792 | |
| | | | | (0.128) | (0.0572) | |
| *Attribute* | | | | | | |
| Surface | 0.108*** | | 0.105*** | | 0.103*** | 0.104*** |
| | (0.0119) | | (0.0120) | | (0.0118) | (0.0118) |
| $Surface^2$ | -0.00064*** | | -0.000628*** | | -0.000614*** | -0.000617*** |
| | (9.33e-05) | | (9.18e-05) | | (9.25e-05) | (9.23e-05) |
| Age | 0.00964*** | | 0.0116*** | | 0.0122*** | 0.0122*** |
| | (0.00280) | | (0.00286) | | (0.00277) | (0.00278) |
| Signature | 0.0400 | | 0.0367 | | 0.0274 | 0.0316 |
| | (0.101) | | (0.103) | | (0.101) | (0.102) |
| Dated | 0.335** | | 0.356** | | 0.325** | 0.319** |
| | (0.150) | | (0.149) | | (0.146) | (0.146) |
| *Other Control* | | | | | | |
| Material | control | | control | | control | control |
| City | control | | control | | control | control |
| Salesroom | control | | control | | control | control |
| Salesyear | control | | control | | control | control |
| Constant | 8.643*** | 6.176*** | 4.242*** | -83.57*** | -45.40*** | -42.73*** |
| | (0.719) | (1.533) | (1.524) | (18.04) | (15.41) | (15.54) |
| Observations | 720 | 720 | 720 | 720 | 720 | 720 |
| R-squared | 0.483 | 0.078 | 0.509 | 0.121 | 0.521 | 0.520 |
| Adj | 0.459 | 0.075 | 0.484 | 0.116 | 0.495 | 0.495 |

Robust standard errors in parentheses. *** p<0.01, ** p<0.05, * p<0.1

Our results show that the two measurements of painting effort, the variances of lines and colors, both significantly positively explain the sales price of the painting at 1% level in most model specifications. It proves our premise that when more painting effort is involved during the production of a painting, the painting will be valued higher. The market indeed priced in the painting effort by the artist.

We also observe that the estimated coefficient of the line variance is generally bigger than that of the color variance. So the line, i.e., the contour of the picture, plays a more important role than the color in determining the sales price for Picasso. It reveals the particular value structure for Picasso paintings. Such features, of course, can vary across different artists from different artistic schools.

The quadratic term of Lline is significantly negatively related to the price, which implies a decreasing marginal effect of line variance to the painting value. However, the square of Lcolor shows as insignificant. We hold that it may be due to the fact that the color variances of Picasso paintings in our sample may not have reached their inflection points to prices yet.

So the potential multicollinearity between Lcolor and its quadratic term explains why they are insignificant once put together in the regression.

Other control variables produce the usual results similar as in the related literature. For the detailed regression outputs that include all control variables, please see Appendix B.

## 4. Applications of the variance measurements in different contexts

We will apply our new variance measurements of lines and colors of paintings to various contexts so as to test their robustness. Also, we hope they can help to reveal more insights concerning the art pricing patterns by the market.

### *4.1 Picasso paintings at different stages*

Our sample includes Picasso's paintings almost across all stages in his entire career. We follow Czujack (1997) and classify Picasso's artworks into eight different periods starting from his childhood to the old age. From Table 3, we can see that the most expensive painting in our sample is Les femmes d'Alger (Version 'O') at the price of 179, 365, 000 USD sold on May 11, 2015. It was produced during the period of Analytical and Synthetic Cubism where Picasso was 33 years old. The cheapest one (Vase tapestry, sold on 2012 at the price of 1280 USD) was produced during the period of Politics and Art in his eighties.

The paintings in his early years are generally more valuable, and the most expensive ones are concentrated in Blue and Rose Period, as well as the period of Analytical and Synthetic Cubism.

**Table 3. Descriptive Statistic of Picasso Paintings at Eight Different Periods（USD）**

| Period | | Obs | Mean | Sd | Min | Max |
|---|---|---|---|---|---|---|
| 1 | Childhood and Youth (1881–1901) | 34 | $6.20\times10^6$ | $1.38\times10^7$ | 75545 | $6.75\times10^7$ |
| 2 | Blue and Rose Period (1902–1906) | 11 | $1.79\times10^7$ | $2.91\times10^7$ | 380000 | $9.30\times10^7$ |
| 3 | Analytical and Synthetic Cubism(1907–1915) | 33 | $1.16\times10^7$ | $3.39\times10^7$ | 88011 | $1.79\times10^8$ |
| 4 | Camera and Classicism (1916–1924) | 72 | $2.94\times10^6$ | $4.76\times10^6$ | 150000 | $3.05\times10^7$ |
| 5 | Juggler of the Form (1925–1936) | 79 | $1.09\times10^7$ | $1.68\times10^7$ | 10586 | $1.06\times10^8$ |
| 6 | Guernica and the "Style Picasso" (1937–1943) | 94 | $7.62\times10^6$ | $1.29\times10^7$ | 160000 | $8.50\times10^7$ |
| 7 | Politics and Art (1944–1953) | 102 | $3.34\times10^6$ | $5.32\times10^6$ | 1280 | $2.99\times10^7$ |
| 8 | The Old Picasso (1954–1973) | 295 | $4.48\times10^6$ | $5.43\times10^6$ | 32567 | $3.69\times10^7$ |

We wish to analyze if the painting effort (i.e., the variations of lines and colors) will have different contributions to the valuation of Picasso's artworks that were produced in his different artistic stages. For this goal, we divide the sample into eight subsamples based on the period classification in Table 3, and run a simple cross-sectional regression that includes only the line and color variances as the regressors for each subsample.

We leave out other control variables because some of the subsamples have very few observations. For example, we only have eleven observations for the second period of Blue and Rose. Hence a regression with all controls may lack the enough degree of freedom. Since the observations within each subsample have very similar attributes, it does not cause a serious problem if we leave out other controls.[9]

From the subsample regression results in Table 4, we can obtain some insights on how the market values different types of artworks of Picasso.

**Table 4. Different Contributions of Painting Efforts at Eight Periods of Picasso**

| | (1) | (2) | (3) | (4) | (5) | (6) | (7) | (8) |
|---|---|---|---|---|---|---|---|---|
| VARIABLES | Lprice | Lprice | Lprice | Lprice | Lprice | Lprice | Lprice | Lprice |
| Lline | 1.218 | -4.51*** | 0.428 | 0.315 | 0.633 | 1.813** | 0.529 | 2.035*** |
| | (1.764) | (1.336) | (2.099) | (0.770) | (0.835) | (0.784) | (0.920) | (0.400) |
| Lcolor | 0.844 | 0.0725 | 0.428 | 0.266 | 1.246*** | 0.344 | 0.933** | 0.328** |
| | (0.535) | (0.620) | (0.540) | (0.267) | (0.234) | (0.327) | (0.417) | (0.160) |
| Constant | 4.575 | 36.14*** | 10.53 | 11.31*** | 5.689 | 5.010 | 6.638 | 3.807** |
| | (9.642) | (6.943) | (10.94) | (3.997) | (3.976) | (3.699) | (4.477) | (1.847) |
| Observations | 34 | 11 | 33 | 72 | 79 | 94 | 102 | 295 |
| R-squared | 0.114 | 0.413 | 0.031 | 0.028 | 0.271 | 0.071 | 0.063 | 0.117 |

Robust standard errors in parentheses. *** $p<0.01$, ** $p<0.05$, * $p<0.1$

We can see that the variances of lines and colors are not significant in general to explain the sales price for the first four periods. The surprising result of the second period that the variance of lines negatively affects the price is primarily due to an outlier in this smallest subsample (only eleven observations). This outlier has a large portion of black background that cracked due to the time. These cracks are considered as lines by our algorithm since our graphic calculation is completely based on its digital image, which consequently overestimates the line variance.[10] The coefficient becomes insignificant once this outlier is eliminated from the subsample.

Overall, the painting effort measurements seem to be unrelated to the extremely high values of these paintings in Picasso's first four periods. In his young ages (Period one to two), most of his traded paintings largely follow the impressionistic trend at the time. The values of such artworks are generally not based on the variations of lines and colors, rather on the so-called artistic impression from the painting.[11]

[9] We also test by introducing more control variables, and the results in Table 4 are very robust.

[10] This painting outlier is *Femme de profil Couverture d'album*, 1904.

[11] We have also tested by running regression (1) for the representative artist of the impressionism, Monet. We find the same results hold, i.e., the variances of lines and colors do not significantly affect the sales price.

In the subsequent Period three to four, Picasso started to develop his own revolutionary styles that deform and reassemble the objects. It is not hard to comprehend that the values of these paintings rest mostly upon the conceptual/notional effort rather than the actual painting effort.

In Picasso's later years, he applied his early invented styles to more extended scenarios, such as personal life, war, and political events, etc. Consequently the conceptual values of these paintings start to decrease, and the painting effort becomes more and more significant in differentiating these artworks. It seems our analysis can provide a quantitative guideline for the valuation of Picasso's dynamic styles throughout his career.

*4.2 Artists across different countries and styles*

We next try to see if our suggested measurements also work for artists from other countries and other painting styles. So we further analyze the sales records for the French impressionism master Pierre-Auguste Renoir, and the traditional Chinese painting master Baishi Qi. We choose them because both are the major representatives for their distinct art schools, and the numbers of their auction records are also large enough for robust inferences.

We obtain 1,476 auction records for Baishi Qi from Artron, and 1,147 records for Pierre-Auguste Renoir from BASI. Please see Appendix B for the summary statistics of these two artists. We only provide their regression outputs in the following Table 5.

**Table 5. Results for Baishi Qi, Pierre-Auguste Renoir and All Three Artists Combined.**

| VARIABLES | (Baishi Qi)<br>Lprice | (Pierre-Auguste Renoir)<br>Lprice | (Three Combined)<br>Lprice |
|---|---|---|---|
| Lline | 0.252** | 0.260 | 0.164* |
| | (0.1101) | (0.170) | (0.0937) |
| Lcolor | 0.184*** | 0.157*** | 0.267*** |
| | (0.0351) | (0.0357) | (0.0271) |
| Surface | 0.258*** | 0.838*** | 0.151*** |
| | (0.0140) | (0.0356) | (0.0119) |
| $Surface^2$ | -0.00367*** | -0.0346*** | -0.000959*** |
| | (0.000211) | (0.00269) | (0.000127) |
| Signature | 0.308*** | 0.464*** | 0.134*** |
| | (0.0714) | (0.0700) | (0.0375) |
| Dated | -- | 0.00628 | 0.300*** |
| | -- | (0.167) | (0.0434) |
| Material | control | control | control |
| City | control | control | control |
| Salesroom | control | control | control |
| Salesyear | control | control | control |
| Painter | | | control |
| Constant | 8.014*** | 8.021*** | 8.118*** |
| | (0.410) | (0.646) | (0.523) |
| Observations | 1,476 | 1,147 | 3,343 |
| R-squared | 0.546 | 0.620 | 0.591 |

Robust standard errors in parentheses. *** p<0.01, ** p<0.05, * p<0.1

We can see that both the variances of lines and colors are significantly positively related to the auction price for Baishi Qi. Here the painting effort does influence the valuation of traditional Chinese portraits. While for Pierre-Auguste Renoir, only the estimated coefficient of the variance of colors is significantly positive. Given the impressionistic nature of his paintings, this result is not surprising. We have known that colors have their own meanings that can evoke corresponding emotions (Valdez and Mehrabian, 1994). Hence they play a crucial role in impressionism for the artist's pictorial expression.[12]

If we further test the combined sample of all three artists, i.e., Pablo Picasso, Pierre-Auguste Renoir and Baishi Qi together, we find that our results are fairly robust: the variations of lines and colors of their paintings are both significantly and positively related to the prices.

*4.3 Cross effect*

Finally, we wish to investigate if the pricing of a painting considers the interaction between the variances of lines and colors, and if the variances of lines and colors also interact with other attributes of the painting.

We carry out the test of cross effect mainly based on the auction data of Picasso, and the following Table 6 provides the results from various model specifications. Both the variances of lines and colors still remain significant, but their cross term is insignificant. It implies that the variations of lines and colors contribute to the pricing of Picasso paintings in a rather independent way.

It is not surprising to see that the variances of lines and colors interact with other painting attributes, notably the size of the painting. The contributions of the variances of lines and colors to the painting price both decrease with the painting size. From this result, we learn that a larger painting may actually value its content complexity even less.

Further robustness analysis with the sample of Chinese traditional paintings by Baishi Qi shows a similar negative relationship between the size and the content complexity (see Table B5 in Appendix B). However, the estimated coefficient of the interactive term between the variances of lines and colors turns out to be significantly positive there. It implies that the variances of lines and colors are mutually reinforcing to their influence on the prices of Chinese traditional paintings.

We understand that the paintings of Picasso and Baishi are of completely different genres. From their summary statistics, we can see that the mean values of variances of lines and colors of Baishi are both smaller than those of Picasso. In general, Chinese traditional paintings have relatively simpler contour and hue composition, because they emphasize more on expression rather than on realism. So the marginal effect of the variations of lines and colors to each other can be significantly increasing.

[12] Notice that the result of Renoir here is different from that of Monet (as noted in footnote 11). Renoir is considered as more classically impressionistic hence the variances of colors of his paintings still significantly affect the sales prices. While Monet is considered as more absolutely impressionistic hence neither variance of lines nor colors is significantly related to the value of the painting.

**Table 6. Cross Effect of the Variances of Lines and Colors for Paintings of Picasso**

| VARIABLES | (1)<br>Lprice | (2)<br>Lprice | (3)<br>Lprice | (4)<br>Lprice | (5)<br>Lprice |
|---|---|---|---|---|---|
| Lline | 0.593** | 0.586** | 0.560* | 0.570** | 0.525* |
| | (0.258) | (0.267) | (0.310) | (0.255) | (0.310) |
| Lcolor | 0.428*** | 0.432*** | 0.425*** | 0.365*** | 0.348*** |
| | (0.0839) | (0.0827) | (0.0832) | (0.0824) | (0.0812) |
| Lline*Lcolor | | -0.0912 | | | 0.205 |
| | | (0.445) | | | (0.435) |
| Surface | 0.0357*** | 0.0357*** | 0.0372*** | 0.311*** | 0.325*** |
| | (0.0120) | (0.0120) | (0.0114) | (0.0583) | (0.0632) |
| Lline*Surface | | | -0.0198 | | -0.0357 |
| | | | (0.0719) | | (0.0660) |
| Lcolor*Surface | | | | -0.0492*** | -0.0513*** |
| | | | | (0.0106) | (0.0116) |
| Age | 0.00563* | 0.00557* | 0.00565* | 0.00789*** | 0.00815*** |
| | (0.00296) | (0.00299) | (0.00294) | (0.00292) | (0.00291) |
| Signature | 0.0376 | 0.0370 | 0.0321 | 0.0437 | 0.0353 |
| | (0.106) | (0.106) | (0.106) | (0.106) | (0.105) |
| Dated | 0.331** | 0.333** | 0.337** | 0.343** | 0.350** |
| | (0.154) | (0.153) | (0.153) | (0.153) | (0.151) |
| Material | Control | control | control | control | Control |
| City | Control | control | control | control | Control |
| Salesrooms | Control | control | control | control | Control |
| Year | Control | control | control | control | Control |
| Constant | 4.478*** | 4.488*** | 4.609*** | 3.251** | 3.413** |
| | (1.438) | (1.454) | (1.667) | (1.444) | (1.667) |
| Observations | 720 | 720 | 720 | 720 | 720 |
| R-squared | 0.436 | 0.436 | 0.436 | 0.455 | 0.457 |

Robust standard errors in parentheses. *** p<0.01, ** p<0.05, * p<0.1

## 5. Conclusion

The above analysis demonstrates that introducing computer graphics into the art pricing research can be valuable. We argue that the creation of a painting calls for a combination of conceptual effort and painting effort from the artist. Consequently, both efforts shall constitute the important price determinants. However, they are two long missing variables in the traditional hedonic model, primarily due to the fact that they are hard to measure.

This paper applies the image recognition techniques to the digital pictures of the auctioned paintings from various famous artists, and calculates the variances of lines and colors that comprise of these paintings. We propose that these variances can act as the good proxies for an artist's painting effort during her composition. So they shall be brought into the hedonic model. We then test their significance in different contexts. Our results prove the robustness of the line and color variances in explaining a painting's market value.

This research for the first time tries to quantify the information content of a painting with tools from computer graphics. The suggested variance measurements can better capture the content heterogeneity of paintings hence improving on the traditional art pricing methodology. Also, our approach can help identify the core value of an artwork (conceptual versus painting) from a quantitative perspective, complementing the usual subjective art appraisals by experts.

As a first attempt to combine computer graphics and art pricing, our suggested measurements still need further refinements in the future studies. For example, besides the line and color variances, we may consider other quantitative indicators that can account for the distance or symmetry of forms for a painting. Furthermore, we can extend our measurements to higher dimensions, e.g., the three-dimension. Then we can also study the pricing of other types of artworks, such as sculptures or even architectures.

## References

Anderson, S C., Ekelund, R B., Jackson, J D, Tollison, R D (2016). Investment in early american art: the impact of transaction costs and no-sales on returns. *J. Cult. Econom*, 40(3), 335-357.

Biey, M L, Zanola, R (2005). The market for picasso prints: a hybrid model approach. *J. Cult. Econom*, 29(2), 127-136.

Buelens, N, Ginsburgh, V (1993). Revisiting Baumol's 'art as a floating crap game'. *Eur. Econom. Rev*. 37, 1351–1374.

Chanel, O (1995). Is art market behavior predictable? *Eur. Econom. Rev*. 39, 519–527.

Etro, F, Stepanova, E (2017). Art collections and taste in the spanish siglo de oro. *J. Cult. Econom*,41:309-335.

Czujack, C (1997). Picasso paintings at auction, 1963-1994. *J. of Cult. Econom,* 21(3), 229-247.

Czujack, C, Martins, M F O (2004). Do art specialists form unbiased pre-sale estimates? an application for picasso paintings. *Appl. Econom. Lett.* 11(4), 245-249.

Forsund, F R, & Zanola, R.(2006).The art of benchmarking: picasso prints and auction house performance. *Appl. Econom.*, 38(12), 1425-1434.

Galenson, D W (2009).*Conceptual revolutions in twentieth-century art*,(Cambridge University Press, New York).

Ginsburgh, V, Radermecker, A-S, Tommasi, D (2019).The effect of experts' opinion on prices of art works: The case of Peter Brueghel the Younger. *J. of Econom. Behavior and Organization*,159,36-50.

Goetzmann, W N (1993). Accounting for taste: art and the financial markets over three centuries. *The Amer. Econom. Rev*, 83(5), 1370-1376.

Korteweg, A, Kräussl, R, Verwijmeren, P (2015). Does it pay to invest in art? A selection-corrected returns perspective. *The Rev. of Financial Studies*, 29(4), 1007-1038.

Lazzaro, E (2006). Assessing quality in cultural goods: the hedonic value of originality in Rembrandts prints. *J. Cult. Econom*, 30(1), 15-40.

Matsuda T, Kitajo K, Yamaguchi Y., & Komaki F (2017). A point process modeling approach for investigating the effect of online brain activity on perceptual switching. *Neuroimage*,152, 50–59.

Mei, J, Moses, M (2002). Art as an investment and the underperformance of masterpieces. *Social Sci. Electronic Publishing*, 92(5), 1656-1668.

Park, H, Ju, L, Liang, T, Tu, Z (2016). Horizon analysis of art investments: evidence from the chinesemarket. *Pacific-Basin Finance Journal*, 41, 17-25.

Pesando, J E (1993). Art as an investment: the market for modern prints. *Amer. Econom. Rev*, 83(5), 1075-1089.

Pesando, J E., & Shum, P M (1999). The returns to Picasso's prints and to traditional financial assets, 1977 to 1996. *J. Cult. Econom*, 23(3), 183-192.

Pesando, J E, Shum, P M, (2007). The law of one price, noise and "irrational exuberance": the auction market for Picasso prints. *J. Cult. Econom.* 31, 263–277.

Renneboog, L, Spaenjers, C (2013). Buying beauty: on prices and returns in the art market. *Manage. Sci.* 59(1), 36-53.

Scorcu, A E, Zanola, R (2011). The “Right” price for art collectibles: a quantile hedonic regression investigation of Picasso paintings. *The J. of Alternative Investments* , 14 (2) 89-99.

Taylor, D,  Coleman, L (2011). Price determinants of aboriginal art, and its role as an alternative asset class. *J. of Banking & Finance*, 35(6), 1519-1529.

Yokoyama,H, Nambu., I, Izawa, J, Wada. Y (2018). Alpha phase synchronization of parietal areas reflects switch-specific activity during mental rotation: an eeg study. *Frontiers in Human Neuroscience,* 12, 259,1-15.

Valdez, P, Mehrabian, A (1994). Effects of color on emotions. *J.of experimental psychology: General*,123(4), 394-409.

## Appendix A

*The calculation of variances of lines and colors for a painting*

The two basic building blocks of a painting are its composing lines and colors. For a more complex painting, it involves more variations of lines and colors. Hence it may cost more effort for the artist to produce. We obtain the digital images of auctioned paintings from various artists, and try to find the numerical measurements for the variations of each painting's lines and colors.

The basic unit of a digital image is called pixel. More pixels in a given area imply a higher resolution of the image. The resolution is determined by the numbers of pixels in both horizontal and vertical directions. In Figure A1, if we enlarge the crossing point of two lines in a picture (Picture A), we can see that they are actually made up of pixels of white and black (Picture B).

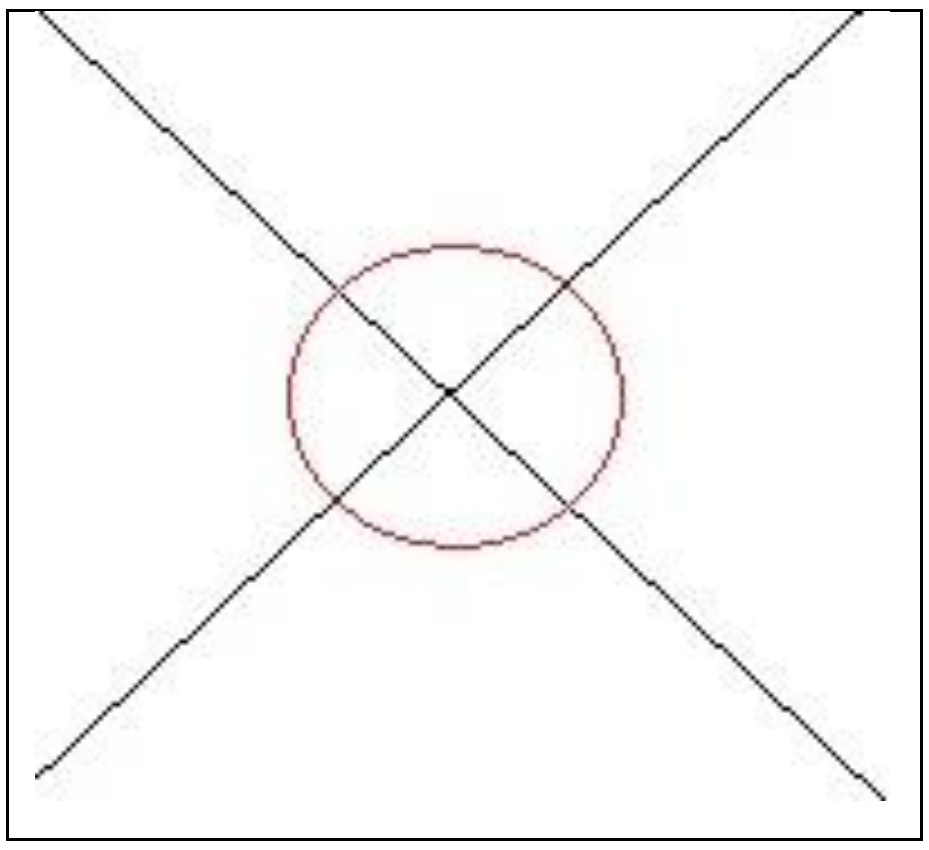

(A)

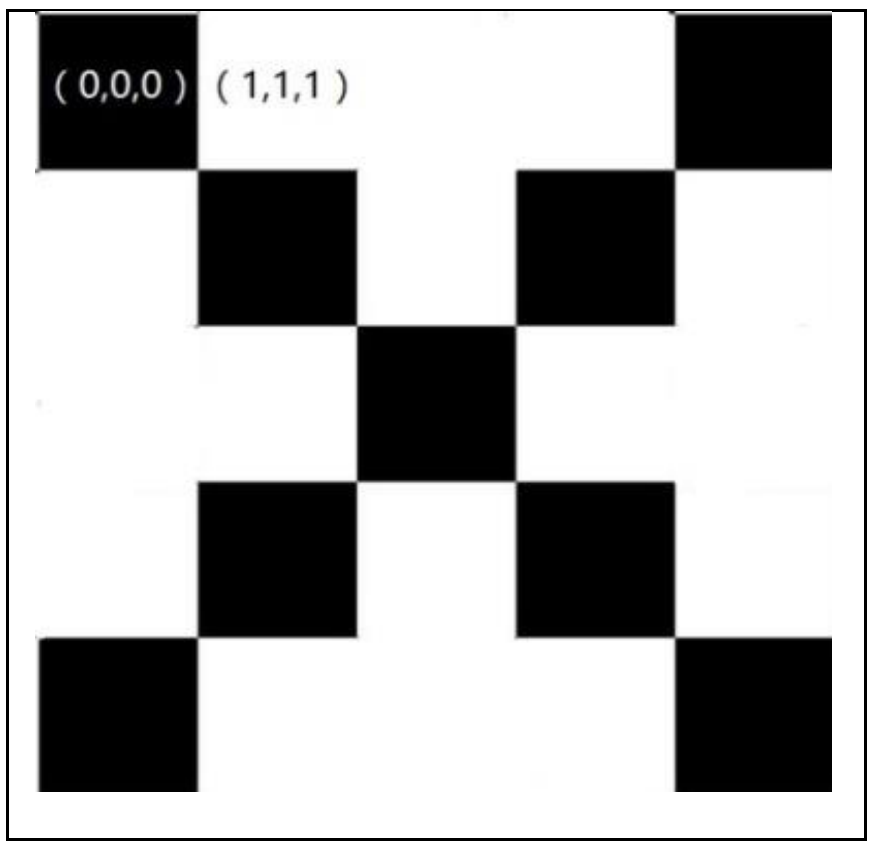

(B)

**Figure A1. Representation of Pixels in a Digital Image**

Each pixel can be characterized by its unique hue and location in an image. The arrays of pixels constitute lines, and the combinations of hues form colors. So our goal is to use the variances of lines and colors of a painting as measurements for the complexity of the painting content hence the painting effort by the artist.

The hue of a pixel can be described by a RGB system, where RGB stands for the proportions of three primary colors, red, green and blue, in a pixel respectively. For example, RGB (0,0,0) means the proportions of red, green and blue are all 0%, so the pixel is black. While, RGB (1,1,1) indicates the proportions of three colors are 100%, then the pixel shows as white.

*I. The variance of lines*

In order to obtain the variance of lines of a painting, we need to follow three steps as illustrated in Figure A2.

First, we convert a colorful picture into a gray one via the floating-point algorithm.[13] In Figure A2, we can see that the black and white Picture B is produced by this algorithm from the original colorful Picture A.

Second, we apply the edge detection method to obtain the image's line structure attribute.[14] Picture C in Figure A2 is generated from Picture B by the edge detection algorithm.

Finally, the variance of lines, $A_i$ in Equation (1), is calculated based on the grayscales of those edges after the edge detection treatment according to the following formula:

$$\frac{\sum_i^N (grayscale_i - \mu)^2}{N} \qquad \text{(A1)}$$

where grayscale$_i$ $\in \{0,1\}$, is the grayscale value of pixel $i$. $\mu$ is the average of all the grayscales of pixels and $N$ is the total number of pixels in the edge detection image of a painting.

[13] The floating-point algorithm calculates a grayscale for a pixel with certain RGB by the formula: grayscale=R*0.3+G*0.59+B*0.11. The grayscale determines a relative distance (or grayness) between white and black, and the color of the pixel is transformed into gray based on this scale.

[14] Edge detection is a basic method in image processing and computer vision, whose purpose is to identify the points with obvious grayscales changes in an image. Edge detection reduces the amount of data by removing the irrelevant information from the image, while retaining the image's important structural attributes.

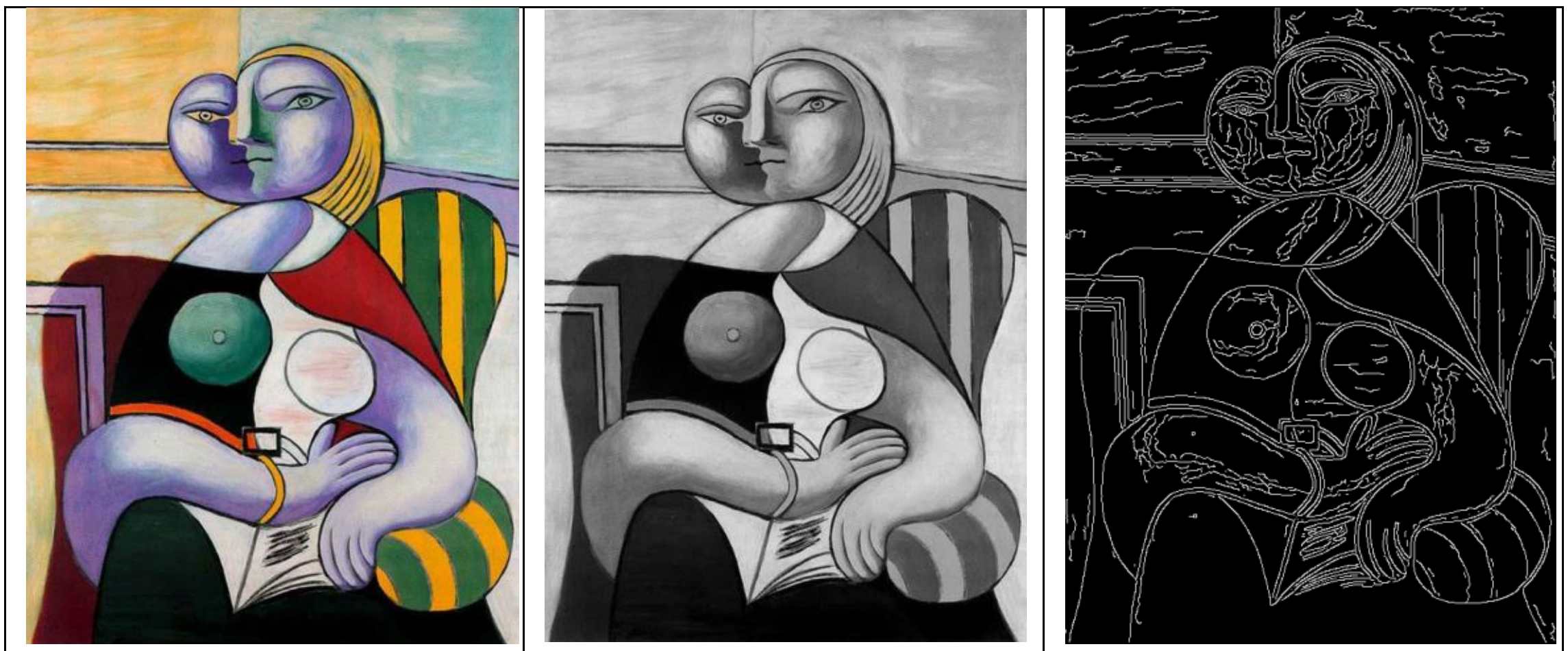

A. Original image | B. Grayscale image | C. Edge detection image

**Figure A2 . Three Steps to Calculate the Variance of Lines**

*II. The variance of colors*

Similarly, the variance of colors, $B_i$ in Equation (1), can be defined as the variance of the hue values of all pixels in the image of a painting. As we describe above, the hue of a pixel can be characterized by a three-dimensional RGB system. While, the hue value is a one-dimensional degree number that is reduced from the RGB to specifically represent the unique color of a given pixel.

The hue value is defined to range from 0 ° to 360 °, starting from the red and go in an anti-clockwise direction. Red is 0 °, green is 120 °, and blue is 240 ° and so on. Let R,G,B be the three proportion numbers in the RGB system, and max and min be the maximum and minimum of R, G, and B, then the hue value of a pixel can be defined by the following formula:

$$hue\ value = \begin{cases} \text{undefined}, & \text{if max} = \text{min} \\ 60° \times \frac{G-B}{max-min} + 0°, & \text{if max} = \text{R and G} \geq \text{B} \\ 60° \times \frac{G-B}{max-min} + 360°, & \text{if max} = \text{R and G} < \text{B} \\ 60° \times \frac{B-R}{max-min} + 120°, & \text{if max} = \text{G} \\ 60° \times \frac{B-R}{max-min} + 120°, & \text{if max} = \text{B} \end{cases} \quad \text{(A2)}$$

Then the variance of hue values can be defined as:

$$\frac{\sum_i^N (huevalue_i - \mu)^2}{N} \quad \text{(A3)}$$

$huevalue_i$ is the hue value of pixel *i* from formula (A2) and $\mu$ is the average of all the hue values of pixels and *N* is the total number of pixels in the painting image.

# Appendix B

*Tables with more detailed outputs*

**Table S1. The Regression Results with all Controls of Table 2**

| VARIABLES | (1) Lprice | (2) Lprice | (3 Lprice | (4) Lprice | (5) Lprice | (6) Lprice |
|---|---|---|---|---|---|---|
| *Effort* | | | | | | |
| Lline | | 1.142*** | 0.537** | 42.34*** | 23.55*** | 23.89*** |
| | | (0.304) | (0.264) | (7.857) | (6.671) | (6.643) |
| $Lline^2$ | | | | -4.606*** | -2.570*** | -2.607*** |
| | | | | (0.879) | (0.738) | (0.735) |
| Lcolor | | 0.624*** | 0.404*** | -0.246 | -0.354 | 0.383*** |
| | | (0.120) | (0.0775) | (1.251) | (0.529) | (0.0783) |
| $Lcolor^2$ | | | | 0.0859 | 0.0792 | |
| | | | | (0.128) | (0.0572) | |
| *Attribute* | | | | | | |
| Surface | 0.108*** | | 0.105*** | | 0.103*** | 0.104*** |
| | (0.0119) | | (0.0120) | | (0.0118) | (0.0118) |
| $Surface^2$ | -0.00064*** | | -0.000628*** | | -0.000614*** | -0.000617*** |
| | (9.33e-05) | | (9.18e-05) | | (9.25e-05) | (9.23e-05) |
| Age | 0.00964*** | | 0.0116*** | | 0.0122*** | 0.0122*** |
| | (0.00280) | | (0.00286) | | (0.00277) | (0.00278) |
| Signature | 0.0400 | | 0.0367 | | 0.0274 | 0.0316 |
| | (0.101) | | (0.103) | | (0.101) | (0.102) |
| Dated | 0.335** | | 0.356** | | 0.325** | 0.319** |
| | (0.150) | | (0.149) | | (0.146) | (0.146) |
| *Other Control Variables* | | | | | | |
| Material | | | | | | |
| Board | 1.912*** | | 1.655*** | | 1.706*** | 1.700*** |
| | (0.574) | | (0.555) | | (0.534) | (0.530) |
| Canvas | 2.157*** | | 1.977*** | | 1.951*** | 1.944*** |
| | (0.538) | | (0.533) | | (0.503) | (0.499) |
| Cardborad | 1.825*** | | 1.650*** | | 1.639*** | 1.622*** |
| | (0.561) | | (0.541) | | (0.513) | (0.511) |
| Panel | 2.120*** | | 1.913*** | | 1.929*** | 1.914*** |
| | (0.555) | | (0.548) | | (0.518) | (0.515) |
| Paper | 1.112** | | 1.042* | | 1.139** | 1.113** |
| | (0.547) | | (0.540) | | (0.519) | (0.516) |
| City | | | | | | |
| New York | 0.750** | | 0.741** | | 0.722** | 0.715** |
| | (0.369) | | (0.366) | | (0.356) | (0.353) |
| London | 0.913** | | 0.919** | | 0.912** | 0.903** |
| | (0.374) | | (0.367) | | (0.358) | (0.355) |
| Paris | 0.0490 | | 0.163 | | 0.235 | 0.228 |
| | (0.392) | | (0.397) | | (0.395) | (0.392) |
| Salesroom | | | | | | |
| Christie's | 0.662*** | | 0.661*** | | 0.689*** | 0.677*** |
| | (0.207) | | (0.206) | | (0.207) | (0.207) |
| Sotheby's | 0.601*** | | 0.639*** | | 0.674*** | 0.653*** |
| | (0.207) | | (0.207) | | (0.207) | (0.207) |
| Saleyear | | | | | | |
| 2001.year | -0.380 | | -0.466 | | -0.281 | -0.295 |
| | (0.377) | | (0.369) | | (0.358) | (0.358) |
| 2002.year | -0.212 | | -0.291 | | -0.322 | -0.341 |
| | (0.373) | | (0.364) | | (0.355) | (0.355) |
| 2003.year | 0.235 | | 0.153 | | 0.136 | 0.126 |
| | (0.386) | | (0.370) | | (0.355) | (0.356) |
| 2004.year | 0.821** | | 0.708** | | 0.632* | 0.636* |
| | (0.365) | | (0.354) | | (0.341) | (0.341) |
| 2005.year | 0.751** | | 0.602* | | 0.544 | 0.536 |
| | (0.359) | | (0.353) | | (0.342) | (0.342) |
| 2006.year | 0.909** | | 0.768** | | 0.671* | 0.670* |
| | (0.373) | | (0.363) | | (0.353) | (0.352) |
| 2007.year | 1.073*** | | 0.930*** | | 0.879*** | 0.880*** |
| | (0.346) | | (0.338) | | (0.325) | (0.325) |
| 2008.year | 1.268*** | | 1.209*** | | 1.208*** | 1.192*** |
| | (0.344) | | (0.335) | | (0.321) | (0.322) |
| 2009.year | 0.671* | | 0.578* | | 0.629* | 0.602* |
| | (0.351) | | (0.342) | | (0.332) | (0.330) |
| 2010.year | 0.902** | | 0.790** | | 0.832** | 0.815** |
| | (0.356) | | (0.350) | | (0.340) | (0.340) |
| 2011.year | 1.360*** | | 1.246*** | | 1.255*** | 1.255*** |
| | (0.353) | | (0.348) | | (0.329) | (0.331) |
| 2012.year | 1.140*** | | 0.991*** | | 0.972*** | 0.964*** |
| | (0.378) | | (0.382) | | (0.365) | (0.365) |
| 2013.year | 1.268*** | | 1.113*** | | 1.078*** | 1.080*** |
| | (0.355) | | (0.350) | | (0.333) | (0.335) |
| 2014.year | 1.266*** | | 1.142*** | | 1.090*** | 1.075*** |
| | (0.336) | | (0.329) | | (0.315) | (0.315) |
| 2015.year | 1.635*** | | 1.506*** | | 1.491*** | 1.477*** |
| | (0.361) | | (0.353) | | (0.339) | (0.339) |
| 2016.year | 0.987** | | 0.898** | | 0.854** | 0.874** |
| | (0.388) | | (0.375) | | (0.358) | (0.357) |
| 2017.year | 1.350*** | | 1.214*** | | 1.181*** | 1.173*** |
| | (0.372) | | (0.361) | | (0.349) | (0.349) |
| 2018.year | 1.648*** | | 1.437*** | | 1.392*** | 1.387*** |
| | (0.352) | | (0.345) | | (0.331) | (0.331) |
| Constant | 8.643*** | 6.176*** | 4.242*** | -83.57*** | -45.40*** | -42.73*** |
| | (0.719) | (1.533) | (1.524) | (18.04) | (15.41) | (15.54) |
| Observations | 720 | 720 | 720 | 720 | 720 | 720 |
| R-squared | 0.483 | 0.078 | 0.509 | 0.121 | 0.521 | 0.520 |

Robust standard errors in parentheses. *** p<0.01, ** p<0.05, * p<0.1

**Table S2. The Regression Results with all Controls of Table 3**

| | (1) | (2) | (3) | (4) | (5) |
|---|---|---|---|---|---|
| VARIABLES | Lprice | Lprice | Lprice | Lprice | Lprice |
| Lline | 0.593** | 0.586** | 0.560* | 0.570** | 0.525* |
| | (0.258) | (0.267) | (0.310) | (0.255) | (0.310) |
| Lcolor | 0.428*** | 0.432*** | 0.425*** | 0.365*** | 0.348*** |
| | (0.0839) | (0.0827) | (0.0832) | (0.0824) | (0.0812) |
| Lline*Lcolor | | -0.0912 | | | 0.205 |
| | | (0.445) | | | (0.435) |
| Surface | 0.0357*** | 0.0357*** | 0.0372*** | 0.311*** | 0.325*** |
| | (0.0120) | (0.0120) | (0.0114) | (0.0583) | (0.0632) |
| Lline*Surface | | | -0.0198 | | -0.0357 |
| | | | (0.0719) | | (0.0660) |
| Lcolor*Surface | | | | -0.0492*** | -0.0513*** |
| | | | | (0.0106) | (0.0116) |
| Age | 0.00563* | 0.00557* | 0.00565* | 0.00789*** | 0.00815*** |
| | (0.00296) | (0.00299) | (0.00294) | (0.00292) | (0.00291) |
| Signature | 0.0376 | 0.0370 | 0.0321 | 0.0437 | 0.0353 |
| | (0.106) | (0.106) | (0.106) | (0.106) | (0.105) |
| Dated | 0.331** | 0.333** | 0.337** | 0.343** | 0.350** |
| | (0.154) | (0.153) | (0.153) | (0.153) | (0.151) |
| Material | | | | | |
| Board | 1.710*** | 1.719*** | 1.727*** | 1.577*** | 1.581*** |
| | (0.511) | (0.508) | (0.510) | (0.500) | (0.502) |
| Canvas | 2.101*** | 2.107*** | 2.119*** | 1.948*** | 1.961*** |
| | (0.479) | (0.479) | (0.478) | (0.469) | (0.472) |
| Cardborad | 1.621*** | 1.628*** | 1.638*** | 1.496*** | 1.506*** |
| | (0.505) | (0.504) | (0.503) | (0.492) | (0.493) |
| Panel | 1.950*** | 1.957*** | 1.968*** | 1.837*** | 1.847*** |
| | (0.508) | (0.506) | (0.507) | (0.498) | (0.500) |
| Paper | 0.968* | 0.982** | 0.995** | 0.863* | 0.876* |
| | (0.501) | (0.499) | (0.495) | (0.490) | (0.486) |
| City | | | | | |
| New York | 0.832** | 0.834** | 0.840** | 0.798** | 0.808** |
| | (0.350) | (0.348) | (0.349) | (0.350) | (0.351) |
| London | 1.034*** | 1.037*** | 1.048*** | 0.985*** | 1.002*** |
| | (0.350) | (0.349) | (0.347) | (0.350) | (0.349) |
| Paris | 0.147 | 0.146 | 0.161 | 0.164 | 0.192 |
| | (0.379) | (0.380) | (0.375) | (0.379) | (0.380) |
| Salesroom | | | | | |
| Christie's | 0.731*** | 0.726*** | 0.727*** | 0.691*** | 0.692*** |
| | (0.214) | (0.214) | (0.215) | (0.215) | (0.216) |
| Sotheby's | 0.685*** | 0.682*** | 0.680*** | 0.676*** | 0.674*** |
| | (0.215) | (0.214) | (0.216) | (0.216) | (0.217) |
| Salesyear | | | | | |
| 2001.year | -0.558 | -0.556 | -0.556 | -0.605 | -0.607 |
| | (0.383) | (0.384) | (0.381) | (0.383) | (0.382) |
| 2002.year | -0.390 | -0.391 | -0.393 | -0.449 | -0.455 |
| | (0.373) | (0.372) | (0.372) | (0.372) | (0.371) |
| 2003.year | 0.0920 | 0.0904 | 0.0911 | 0.0994 | 0.102 |
| | (0.387) | (0.386) | (0.386) | (0.381) | (0.378) |
| 2004.year | 0.677* | 0.674* | 0.676* | 0.678* | 0.684* |
| | (0.369) | (0.369) | (0.369) | (0.367) | (0.364) |
| 2005.year | 0.354 | 0.356 | 0.356 | 0.370 | 0.371 |
| | (0.375) | (0.376) | (0.376) | (0.369) | (0.371) |
| 2006.year | 0.784** | 0.782** | 0.783** | 0.819** | 0.824** |
| | (0.369) | (0.369) | (0.371) | (0.368) | (0.367) |
| 2007.year | 0.833** | 0.833** | 0.839** | 0.828** | 0.839** |
| | (0.349) | (0.349) | (0.348) | (0.350) | (0.347) |
| 2008.year | 1.153*** | 1.149*** | 1.154*** | 1.155*** | 1.167*** |
| | (0.342) | (0.339) | (0.341) | (0.341) | (0.337) |
| 2009.year | 0.566 | 0.565 | 0.578 | 0.497 | 0.519 |
| | (0.371) | (0.370) | (0.373) | (0.366) | (0.365) |
| 2010.year | 0.745** | 0.747** | 0.752** | 0.699* | 0.707* |
| | (0.365) | (0.366) | (0.365) | (0.365) | (0.366) |
| 2011.year | 1.184*** | 1.185*** | 1.189*** | 1.154*** | 1.159*** |
| | (0.359) | (0.361) | (0.359) | (0.356) | (0.356) |
| 2012.year | 0.951** | 0.949** | 0.944** | 0.941** | 0.933** |
| | (0.371) | (0.369) | (0.369) | (0.370) | (0.367) |
| 2013.year | 1.045*** | 1.046*** | 1.046*** | 1.013*** | 1.012*** |
| | (0.364) | (0.365) | (0.363) | (0.365) | (0.364) |
| 2014.year | 1.132*** | 1.132*** | 1.132*** | 0.985*** | 0.979*** |
| | (0.335) | (0.335) | (0.335) | (0.338) | (0.338) |
| 2015.year | 1.428*** | 1.426*** | 1.428*** | 1.411*** | 1.414*** |
| | (0.370) | (0.369) | (0.369) | (0.368) | (0.365) |
| 2016.year | 0.853** | 0.857** | 0.860** | 0.847** | 0.851** |
| | (0.382) | (0.385) | (0.382) | (0.384) | (0.385) |
| 2017.year | 1.139*** | 1.139*** | 1.142*** | 1.136*** | 1.142*** |
| | (0.371) | (0.372) | (0.370) | (0.370) | (0.367) |
| 2018.year | 1.364*** | 1.364*** | 1.366*** | 1.322*** | 1.322*** |
| | (0.355) | (0.355) | (0.354) | (0.355) | (0.353) |
| Constant | 4.478*** | 4.488*** | 4.609*** | 3.251** | 3.413** |
| | (1.438) | (1.454) | (1.667) | (1.444) | (1.667) |
| Observations | 720 | 720 | 720 | 720 | 720 |
| R-squared | 0.436 | 0.436 | 0.436 | 0.455 | 0.457 |

Robust standard errors in parentheses. *** p<0.01, ** p<0.05, * p<0.1

**Table S3. The Regression Results with all Controls of Table 4**

| | (Baishi Qi) | (Pierre-Auguste Renoir) | (Three Combined) |
|---|---|---|---|
| VARIABLES | Lprice | Lprice | Lprice |
| Lline | 0.252** | 0.260 | 0.164* |
| | (0.1101) | (0.170) | (0.0937) |
| Lcolor | 0.184*** | 0.157*** | 0.267*** |
| | (0.0351) | (0.0357) | (0.0271) |
| Surface | 0.258*** | 0.838*** | 0.151*** |
| | (0.0140) | (0.0356) | (0.0119) |
| $Surface^2$ | -0.00367*** | -0.0346*** | -0.000959*** |
| | (0.000211) | (0.00269) | (0.000127) |
| Signature | 0.308*** | 0.464*** | 0.134*** |
| | (0.0714) | (0.0700) | (0.0375) |
| Dated | -- | 0.00628 | 0.300*** |
| | -- | (0.167) | (0.0434) |
| Material | | | |
| Canvas | paper | 0.690** | 0.0717 |
| | | (0.315) | (0.188) |
| Panel | paper | 0.0429 | -0.252 |
| | | (0.376) | (0.243) |
| Paper | paper | 0.326 | -0.690** |
| | | (1.349) | (0.270) |
| City | | | |
| New York | -0.568*** | 0.623*** | 1.580*** |
| | (0.208) | (0.122) | (0.138) |
| London | -- | 0.464*** | 1.359*** |
| | -- | (0.117) | (0.139) |
| Paris | -- | 0.271** | 0.443*** |
| | -- | (0.110) | (0.140) |
| Hong Kong | -0.191 | -- | 1.328*** |
| | (0.196) | -- | (0.189) |
| Salesroom | | | |
| Christie’s | -0.246 | 0.196** | -0.466*** |
| | (0.321) | (0.0848) | (0.0667) |
| Sotheby’s | -0.331 | 0.133 | -0.680*** |
| | (0.312) | (0.0935) | (0.0736) |
| Salesyear | | | |
| 2001.year | 0.273 | -0.201 | -0.226 |
| | (0.320) | (0.157) | (0.173) |
| 2002.year | 0.596** | -0.0204 | -0.120 |
| | (0.264) | (0.172) | (0.172) |
| 2003.year | 0.791*** | 0.152 | 0.158 |
| | (0.249) | (0.147) | (0.165) |
| 2004.year | 1.390*** | 0.383** | 0.497*** |
| | (0.236) | (0.158) | (0.150) |
| 2005.year | 2.179*** | 0.657*** | 0.690*** |
| | (0.502) | (0.134) | (0.153) |
| 2006.year | 1.429*** | 0.868*** | 0.472*** |
| | (0.216) | (0.129) | (0.144) |
| 2007.year | 1.804*** | 0.828*** | 0.948*** |
| | (0.231) | (0.139) | (0.145) |
| 2008.year | 1.488*** | 0.809*** | 0.855*** |
| | (0.232) | (0.149) | (0.151) |
| 2009.year | 2.332*** | 0.390*** | 0.653*** |
| | (0.241) | (0.132) | (0.150) |
| 2010.year | 2.832*** | 0.624*** | 1.151*** |
| | (0.223) | (0.138) | (0.143) |
| 2011.year | 3.166*** | 0.614*** | 1.558*** |
| | (0.214) | (0.152) | (0.138) |
| 2012.year | 2.713*** | 0.610*** | 1.154*** |
| | (0.218) | (0.142) | (0.143) |
| 2013.year | 2.460*** | 0.605*** | 1.041*** |
| | (0.227) | (0.144) | (0.139) |
| 2014.year | 2.467*** | 0.816*** | 1.124*** |
| | (0.223) | (0.125) | (0.136) |
| 2015.year | 2.808*** | 0.666*** | 1.167*** |
| | (0.229) | (0.126) | (0.142) |
| 2016.year | 2.497*** | 0.445*** | 0.956*** |
| | (0.232) | (0.128) | (0.146) |
| 2017.year | 2.246*** | 0.433*** | 0.839*** |
| | (0.224) | (0.167) | (0.144) |
| 2018.year | 2.211*** | 0.528*** | 1.128*** |
| | (0.235) | (0.156) | (0.149) |
| Painter | | | |
| Picasso | -- | -- | 1.614*** |
| | -- | -- | (0.231) |
| Renoir | -- | -- | 0.494** |
| | -- | -- | (0.244) |
| Constant | 7.995*** | 8.021*** | 8.118*** |
| | (0.405) | (0.646) | (0.523) |
| Observations | 1,476 | 1,147 | 3,343 |
| R-squared | 0.547 | 0.620 | 0.591 |

Robust standard errors in parentheses. *** p<0.01, ** p<0.05, * p<0.1

**Table S4. Summary Statistics of Chinese Artist Baishi Qi**

| VARIABLES | N | Mean | Sd | Min | Max |
|---|---|---|---|---|---|
| Price($) | 1,476 | 544,442 | 2.46E+06 | 1,380 | 6.38E+07 |
| Line | 1,476 | 0.0732 | 0.0205 | 0.0203 | 0.187 |
| Color | 1,476 | 0.151 | 0.0884 | 0.00898 | 0.434 |
| Salesyear | 1,476 | 2,012 | 4.088 | 2,000 | 2,018 |
| Surface(1000cm$^2$) | 1,476 | 3,502 | 2,845 | 138 | 66,025 |
| Signature | 1,476 | 0.476 | 0.5 | 0 | 1 |
| City | | | | | |
| Hongkong | 1,476 | 0.934 | 0.062 | 0 | 0 |
| New York | 1,476 | 0.042 | 0.04 | 0 | 0 |
| London | 1,476 | 0.018 | 0.018 | 0 | 0 |
| Others | 1,476 | 0.006 | 0.006 | 0 | 0 |
| Salesroom | | | | | |
| Christie’s | 1,476 | 0.35 | 0.227 | 0 | 0 |
| Sotheby’s | 1,476 | 0.394 | 0.239 | 0 | 0 |
| Others | 1,476 | 0.256 | 0.191 | 0 | 0 |

**Table S5. Summary Statistics of  the French Artist Pierre-Auguste Renoir**

| VARIABLES | N | Mean | Sd | Min | Max |
|---|---|---|---|---|---|
| Price($) | 1,147 | 675,077 | 1.49E+06 | 900 | 2.10E+07 |
| Line | 1,147 | 0.118 | 0.0197 | 0.0446 | 0.179 |
| Color | 1,147 | 0.132 | 0.0764 | 0 | 0.447 |
| Age | 1,147 | 121.5 | 11.24 | 28 | 161 |
| Salesyear | 1,147 | 2,009 | 5.305 | 2,000 | 2,018 |
| Surface(1000cm$^2$) | 1,147 | 1,284 | 1,512 | 25.81 | 22,645 |
| Signature | 1,147 | 0.619 | 0.486 | 0 | 1 |
| Dated | 1,147 | 0.0444 | 0.206 | 0 | 1 |
| Material | | | | | |
| Canvas | 1,147 | 0.972125 | 0.027098 | 0 | 1 |
| Others | 1,147 | 0.027875 | 0.027098 | 0 | 1 |
| City | | | | | |
| London | 1,147 | 0.347561 | 0.226762 | 0 | 1 |
| New York | 1,147 | 0.391115 | 0.238144 | 0 | 1 |
| Paris | 1,147 | 0.114983 | 0.101762 | 0 | 1 |
| Others | 1,147 | 0.146341 | 0.124926 | 0 | 1 |
| Salesroom | | | | | |
| Christie’s | 1,147 | 0.420732 | 0.243717 | 0 | 1 |
| Sotheby’s | 1,147 | 0.398955 | 0.23979 | 0 | 1 |
| Others | 1,147 | 0.180314 | 0.147801 | 0 | 1 |

**Table S6.  The Regression Results of Cross Effect from Baishi Qi**

| | (1) | (2) | (3) |
|---|---|---|---|
| VARIABLES | Lprice | Lprice | Lprice |
| Lline | 0.412*** | 0.260** | 0.278** |
| | (0.111) | (0.117) | (0.119) |
| Lcolor | 0.179*** | 0.189*** | 0.179*** |
| | (0.0348) | (0.0366) | (0.0356) |
| Surface | 0.000136*** | 0.000154*** | 0.000138*** |
| | (4.62e-05) | (3.89e-05) | (4.68e-05) |
| Lline*Lcolor | 0.503*** | | |
| | (0.164) | | |
| Lline*Surface | | -0.000130* | |
| | | (7.54e-05) | |
| Lcolor*Surface | | | 2.08e-05 |
| | | | (2.56e-05) |
| Signature | 0.297*** | 0.295*** | 0.304*** |
| | (0.0705) | (0.0711) | (0.0714) |
| City | | | |
| New York | -0.579*** | -0.556*** | -0.609*** |
| | (0.221) | (0.212) | (0.220) |
| Hong Kong | -0.151 | -0.140 | -0.165 |
| | (0.199) | (0.191) | (0.195) |
| Singapore | 1.201*** | 1.383*** | 1.287*** |
| | (0.375) | (0.378) | (0.372) |
| Salesroom | | | |
| Christie's | -0.264 | -0.243 | -0.229 |
| | (0.312) | (0.311) | (0.323) |
| Sotheby's | -0.458 | -0.428 | -0.432 |
| | (0.305) | (0.306) | (0.316) |
| Salesyear | | | |
| 2001.year | 0.233 | 0.274 | 0.225 |
| | (0.337) | (0.368) | (0.344) |
| 2002.year | 0.519* | 0.575* | 0.551* |
| | (0.294) | (0.306) | (0.291) |
| 2003.year | 0.697** | 0.836*** | 0.791*** |
| | (0.274) | (0.284) | (0.273) |
| 2004.year | 1.374*** | 1.457*** | 1.427*** |
| | (0.274) | (0.276) | (0.269) |
| 2005.year | 2.132*** | 2.181*** | 2.160*** |
| | (0.419) | (0.449) | (0.447) |
| 2006.year | 1.312*** | 1.386*** | 1.349*** |
| | (0.252) | (0.257) | (0.251) |
| 2007.year | 1.845*** | 1.960*** | 1.898*** |
| | (0.267) | (0.285) | (0.269) |
| 2008.year | 1.478*** | 1.575*** | 1.509*** |
| | (0.265) | (0.275) | (0.265) |
| 2009.year | 2.239*** | 2.350*** | 2.309*** |
| | (0.272) | (0.278) | (0.269) |
| 2010.year | 2.835*** | 2.917*** | 2.878*** |
| | (0.254) | (0.264) | (0.254) |
| 2011.year | 3.085*** | 3.180*** | 3.150*** |
| | (0.246) | (0.254) | (0.244) |
| 2012.year | 2.637*** | 2.732*** | 2.690*** |
| | (0.250) | (0.259) | (0.249) |
| 2013.year | 2.416*** | 2.504*** | 2.461*** |
| | (0.257) | (0.262) | (0.254) |
| 2014.year | 2.426*** | 2.508*** | 2.469*** |
| | (0.254) | (0.261) | (0.252) |
| 2015.year | 2.761*** | 2.869*** | 2.813*** |
| | (0.262) | (0.271) | (0.261) |
| 2016.year | 2.464*** | 2.536*** | 2.508*** |
| | (0.262) | (0.271) | (0.262) |
| 2017.year | 2.216*** | 2.290*** | 2.255*** |
| | (0.255) | (0.266) | (0.255) |
| 2018.year | 2.221*** | 2.287*** | 2.261*** |
| | (0.268) | (0.276) | (0.267) |
| Constant | 9.931*** | 9.827*** | 9.871*** |
| | (0.377) | (0.385) | (0.379) |
| Observations | 1,476 | 1,476 | 1,476 |
| R-squared | 0.507 | 0.506 | 0.502 |

Robust standard errors in parentheses. *** p<0.01, ** p<0.05, * p<0.1